# Modeling Battery Electric Vehicle Users' Charging Decisions in Scenarios with Both Time-Related and Distance-Related Anxiety

Jiyao Wang[1], Wenbo Zhang[1], Xiao (Luke) Wen[2], Dengbo He[1,3] [*], Ran Tu[4]

1. Intelligent Transportation Thrust, The Hong Kong University of Science and Technology (Guangzhou), Guangzhou, China
2. Interdisciplinary Programs Office, The Hong Kong University of Science and Technology, Hong Kong SAR, China
3. Department of Civil and Environmental Engineering, The Hong Kong University of Science and Technology, Hong Kong SAR, China
4. School of Transportation, Southeast University, Nanjing, China

## Abstract

As one of the most promising alternatives to internal combustion engine vehicles, battery electric vehicles (BEVs) have become increasingly prevalent in recent years. However, range anxiety is still a major concern among BEV users or potential users in recent years. The social-psychological factors were found to be associated with range anxiety, but how the charging decisions are affected by range anxiety is still unclear. Thus, in our study, through an online questionnaire issued in mainland China, we collected 230 participants' charging decisions in 60 range-anxiety-inducing scenarios in which both distance-related, and time-related anxiety co-existed. Then, an interpretable machine learning (ML) approach with the Shapley Additive Explanations method was used to model BEV users' charging decisions in these scenarios. To further explore users' decision-making mechanisms, a Bayesian-Network-regression mixed approach was used to model the inner topological structure among the factors influencing users' decisions. We find that both time-related and distance-related factors can affect users' charging decisions, but the influence of waiting time is softer compared to the BEV range. Users' charging decisions can also be moderated by users' psychological states (i.e., range anxiety level and trust in range estimation system), individual differences (i.e., age and personality), and BEV using experience (i.e., driving mileage, display mileage and range estimation cycle of range estimation system), of which, the range anxiety level is more directly related with users' charging decisions. Findings from this study can provide insights into the optimization of charge station distribution and customization of the charging recommendation system.

# Introduction

With the development of electrified vehicle power systems, electric vehicles (EVs) are anticipated to occupy a significant portion of the vehicle market in the near future. The 2020 annual sales report of China (Phyllis Zuo, 2022) predicted that in 2035, over half of all vehicle sales will be new energy vehicles, with battery electric vehicles (BEVs) making up more than 95% of the market share. However, range anxiety is still one of the major obstacles to users' adoption of BEVs. Traditionally, range anxiety manifested as drivers' uncertainty about whether their BEVs have sufficient battery to reach a destination (Rauh et al., 2015). Traditionally, range anxiety was framed and alleviated from the 'distance' perspective (i.e., whether I can reach the destination with the remaining battery) (Rainieri et al., 2023; Xu et al., 2024). As a psychological state, range anxiety can result in cognitive, emotional, behavioral, and physiological changes before the purchase or during the usage of BEVs. For example, range anxiety can negatively affect the likelihood of buying a limited-range BEV (Rauh et al., 2015), and drivers with higher levels of range anxiety have been found to be more active in searching for charge opportunities on a trip (Tian et al., 2023). Thus, it becomes necessary to understand the factors influencing range anxiety and alleviate range anxiety among BEV users or potential users.

Traditionally, range anxiety was framed from the 'distance' perspective (i.e., whether I can reach the destination with the remaining battery) (Rainieri et al., 2023; Xu et al., 2024). However, with the development of charging infrastructure and BEV technologies, reaching a destination may not be problematic in most scenarios; but the time cost for charging might become a greater concern. In fact, despite the fact that many rest areas along highways in the eastern region of China have installed charging stations (The Beijing News, 2022), the queuing time for charging a BEV can take up to four hours during holidays (News China, 2022). This highlights a new dimension of range anxiety: the time cost might also be associated with range anxiety. Our preliminary work (Wang et al., 2023) defined time-related range anxiety as "*a driver's uncertainty about reaching the destination in time.*" Based on a survey study, we confirmed the existence of time-related range anxiety (i.e., driver's uncertainty about reaching the destination in time) and observed the trade-off between distance-related and time-related range anxieties. Thus, the time-related range anxiety should be considered for drivers' charging decision modeling and prediction.

At the same time, previous research (e.g., Noel et al., 2019, Wang et al., 2023) mostly focused on the outcome of range anxiety with linear regression models (e.g., discrete choice modeling (Greene, 2009)). Constrained by the assumptions of statistical models (Wen et al., 2021), statistical methods may not reach satisfactory prediction accuracy and become infeasible when non-linearity exists in the dataset (Chen et al., 2021). Further, in the statistical analysis, correlated factors have to be abandoned or aggregated to avoid collinearity and thus may not fully reveal the hierarchical influence mechanisms leading to a charging decision (Wang et al., 2024). For example, although we have found that both heterogeneous range anxiety and other socio-psychological factors (e.g., trust, driving experience) can affect charging behaviors (Wang et al., 2023), the associations among the factors are still unclear (e.g., whether trust is directly associated with behaviors or by moderating other psychological factors like range anxiety). Thus, previous linear-regression-based work provided limited guidance on the design of the charging recommendation systems and charging station planning.

To overcome the limitations of previous studies and quantify the factors that influence BEV users' charging decisions, in our current study, more diverse scenarios were designed, and more samples were collected on top of (Wang et al., 2023). Further, a mixed approach combining machine learning (ML), Shapley Additive Explanations (SHAP), Bayesian Network (BN) (Friedman et al., 1997), and regression analysis was utilized. First, multiple ML methods were assessed based on the K-fold validation, and the one with the best performance was selected. ML models were adopted as they have the potential to fit complex data structures with higher accuracy compared to regression models. However, the ML models are less interpretable compared to regression models (Seto et al., 2022). Given that we need to reveal the mechanisms leading to the charging decision, two additional analyses were conducted, i.e., the SHAP and BN-regression, to provide a more complete picture of how a charging decision can be affected by range anxiety. The outcomes of SHAP and BN-regression are mutually supplemental. The SHAP method (Lundberg and Lee, 2017) can rank the importance of the influential factor (e.g., trust, driving experience (Wang et al., 2023)) but may not reveal the internal relationships among the factors of the charging decision. To explore the hierarchical relationships among variables of interest, traditional statistical methods such as the Structure Equation Model (SEM) (Pan et al., 2019) were adopted, which cannot model the nonlinear relationship among variables and have hard constraints on the data types being modeled that limit the types of questions used in a questionnaire. Inspired by this, the BN-regression has been proposed to identify topological structures among influential factors and quantify the potential linear relationships along with each edge in BN (Wang et al., 2024).

In all, the main contributions of this study are summarized as follows: (1) To the best of our knowledge, this study, including the part previously reported (Wang et al., 2023), is the first attempt to evaluate how both distance-related and time-related factors can affect BEV users' charging decisions. We are also the first to model how the charging decisions can be moderated by other scenario-related and social-psychological factors; (2) Being different from Wang et al. (2023), we replaced the regression analysis with the ML approach to model charging behaviors so that the extracted model can be more accurate, as ML models do not have multicollinearity problems and thus all influential factors can be kept for BN analysis. This is also the first time that the ML-SHAP-BN-regression mixed approach was utilized to investigate the mechanisms of a decision-making mechanism based on questionnaire data; (3) This study would provide more insights into the association between influential factors of a charging decision on top of Wang et al. (2023). The range anxiety model and users' charging decision margins extracted in this study can also support the optimization of charging station distributions and allow customized BEV driver energy replenishment suggestions by modeling the cognitive mechanisms leading to range anxiety (Wang, Huang, et al., 2024). For example, when optimizing the charging station network, the planners should consider efficiency on top of reachability if time-related anxiety matters, which may help relieve range anxiety and facilitate users' BEV purchasing decisions.

## Related Work

### *Definition and Causes of Range Anxiety*

The concept of range anxiety among BEV drivers was first proposed in 1997 (Nilsson, 2011) and has started to attract increasingly more attention in recent years. In general, range anxiety refers to the concern of not having enough battery range to complete a journey in the BEVs (Chen et al., 2021).

Specifically, from the scenario perspective, King et al. (2015) defined range anxiety as a situation where a driver must travel a distance greater than the typical range of the EV on a single charge. This definition placed a primary emphasis on the "range" component (Neubauer & Wood, 2014). Later, researchers started to realize that range anxiety can also be a psychological status and formally defined range anxiety as "*a psychological reaction to the tension-inducing circumstance of the battery nearing depletion*" (Franke & Krems, 2013). This definition focused more on the "anxiety" component.

To explain why range anxiety exists, Rauh et al. (2015) proposed an influential factor model. They found that prior experience with BEVs could mitigate range anxiety, potentially because users' familiarity with similar range-critical situations enables them to come up with more solutions to handle the current situation, thereby diminishing their perceived range uncertainty. Further, the limited availability of charging stations (Zheng, 2021) and larger "comfort range" (i.e., the remaining range or battery level of a battery electric vehicle (BEV) prior to embarking on a journey) were found to be associated with increased range anxiety (Yuan et al., 2018). At the same time, the system design of BEVs can also influence users' range anxiety. Specifically, as the range estimation system (RES) can integrate and analyze the battery-related parameters and provide an estimation of the battery level, the RES can be regarded as a type of automation. Thus, users' trust in and reliance on the RES can also play a role (Hariharan et al., 2022; Wang et al., 2021).

It should be noted that, in the studies mentioned above, range anxiety was framed only from the perspective of travel distance. However, time uncertainty may also matter to BEV users (Wang et al., 2023). For example, Zhang et al., (2021) found that the time cost of BEV charging can affect customers' willingness to purchase BEVs. As a preliminary study (Wang et al., 2023), we also proposed a revised range anxiety model that takes both distance- and time-related factors into consideration. In the study, the influence of individual heterogeneity on range anxiety was evaluated through a questionnaire issued in mainland China. However, none of the studies mentioned above modeled how range anxiety develops when both distance and time pressure exist.

*Alleviation of Range Anxiety*

Efforts have been made to alleviate range anxiety, mostly from infrastructure optimization or from technology improvement perspectives of view. In recent years, researchers and vehicle manufacturers have always been trying to increase the battery capacity (e.g., Hanifah et al., 2015) and the charging speed of BEVs (e.g., Chakraborty et al., 2022). However, technology development can be slow and unpredictable, and the hardware upgrades of BEV and infrastructure can be costly (Madina et al., 2016). At this stage, although battery technology has developed rapidly, current BEVs remain limited by their batteries' size, weight, and cost (Ullah et al., 2021; Westin et al., 2018). Thus, researchers also tried to tackle the range anxiety issue from the charging station planning perspective of view. For example, solutions have been proposed to improve the charging experience by optimizing the distribution of charging station locations (Bulut & Kisacikoglu, 2017; Hafez & Bhattacharya, 2017; Pan et al., 2020) and by reducing waiting time for queueing (Antoun et al., 2021). However, as human users usually deviate from the rationality (Jones, 1999), BEV users would always expect more charging opportunities. This may conflict with the goal of minimizing capital budget in most optimization procedures. As a result, the equilibrium point reached in previous optimization solutions may not satisfy actual users. Further, although most previous optimization methods included reducing waiting time in the model evaluation (Uslu & Kaya, 2021), BEV users may not necessarily weigh the time pressure and distance pressure equally. Considering that range anxiety has been found to influence users' charging decisions, to better optimize the charging station distribution, it becomes necessary to quantify the relationship between social-psychological factors of range anxiety and charging decisions.

Thus, in recent years, researchers also started to explore how to alleviate users' range anxiety from psychology perspective of view, for example, by providing drivers with more accurate range predictions or by issuing warnings in advance before the remaining battery level becomes too low (Modi et al., 2020; Wang et al., 2018). Some studies even tried to customize the charging recommendations to provide adaptive charging advice with the goal of satisfying individual preferences (Wang et al., 2020; Zhang et al., 2022). At the same time, driver education has also been found to be an effective countermeasure to range anxiety. For example, when drivers are informed of effective eco-driving techniques and available charging options, their range anxiety can be reduced (Hardman & Tal, 2018).

However, again, none of the studies mentioned above considered time-related range anxiety, which has been found to be an important component of the range anxiety (Wang et al., 2023).

*Modeling of BEV Charging Behaviors*

To quantify the impact of range anxiety on BEV adoption and to optimize the distribution of charging infrastructure, it is crucial to model drivers' charging patterns. In general, previous studies can be categorized into two major streams based on their research objectives, i.e., charging behavior prediction and behavioral pattern analysis. For charging behavior prediction, studies usually aim at predicting BEV users' charging behaviors based on historical data with ML methods (e.g., LightGBM). The targeted behaviors include the charging session duration (Ai et al., 2018; Frendo et al., 2020; Xiong et al., 2017) and the charging station choices (Ullah et al., 2023; Yang et al., 2020). Such research usually has a high demand on the sample size of the behavior data (Shahriar et al., 2020). Further, the implicit of the learned knowledge in the ML models makes it difficult for the models to be expanded for the objectives that are beyond the scope of the studies (e.g., a model built for charging station location optimization can hardly be used for charging route planning).

To overcome the limitations of ML-based charging behavior prediction, in recent years, researchers also focused on BEV users' charging choice analysis (Baresch & Moser, 2019; Jin et al., 2013; Morrissey et al., 2016) with statistical regression models. Overall, previous research has explored BEV charging behavior and adoption from the perspectives of activity-based travel patterns, charging infrastructure planning, and traveler preferences. For example, previous work found that approximately 80% of BEV charging events occur at home, while the rest of the charging events happen at the workplace and public locations (Baresch & Moser, 2019; Lee et al., 2020). In another study, Dong et al. (2014) found that BEV users prefer to charge their BEVs at destinations when the vehicles are parked for an extended period of time. However, although previous research attempted to model BEV charging patterns (Wang et al., 2022), few studies considered the influence of psychological factors on users' charging decisions.

**Table 1.** List of Abbreviations

| | |
|---|---|
| BEV | Battery Electric Vehicle |
| ML | Machine Learning |
| SHAP | Shapley Additive Explanations |
| BN | Bayesian Network |
| LR | Logistic Regression |
| Adaboost | Adaptive Boost |
| DT | Decision Tree |
| RF | Random Forest |
| MLP | Multi-Layered Perception |
| RFE | Recursive Feature Elimination |
| DAG | Directed Acyclic Graph |

# Method

In this section, we introduce the adopted methodology framework (see Figure 1) and all abbreviations used in this work are summarized in Table 1. In general, we designed a questionnaire (step one in Figure 1) to collect BEV drivers' scenario-based charging choices. Further, based on the results of our preliminary study (Wang et al., 2023), selected social-psychological characteristics of BEV users were also collected. Then, to model BEV users' charging decisions and extract the decision margins of drivers, ML models were built and selected. Based on the ML model with the best performance, to comprehensively and structurally analyze the influential factors of BEV users' charging decisions, SHAP and BN-regression analyses were conducted. It should be noted that, the BN analysis was used to further investigate the influence of socio-psychology factors on BEV drivers' time-related and distance-related anxiety. Thus, only the hierarchical relationships among scenario-free factors (i.e., socio-psychology variables and drivers' distance- and time-anxiety variables) were modeled in BN-regression analyses.

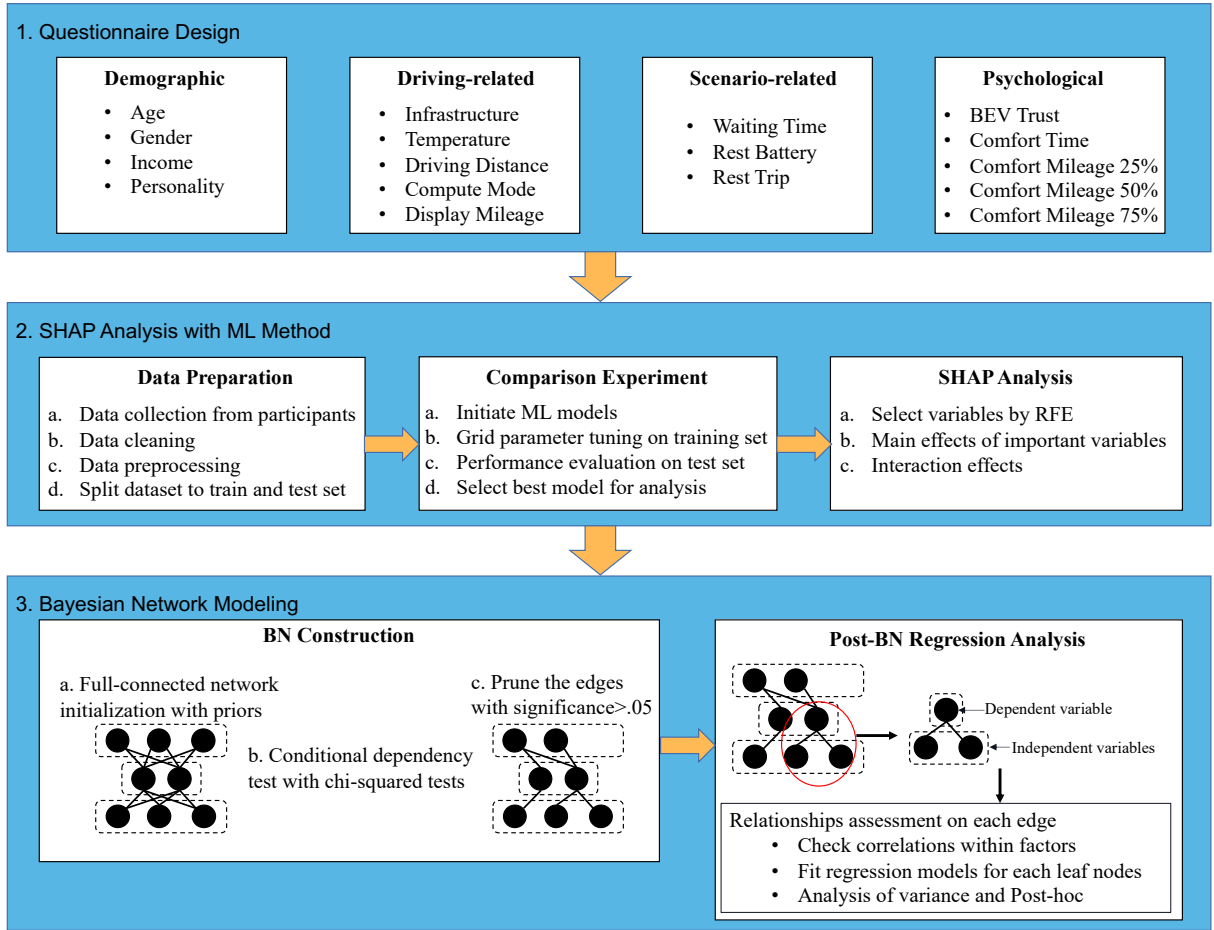

**Figure 1.** The overall methodological framework of this work

*Questionnaire Design*

Table 2 summarizes the questionnaire design in this study.

- The first ten questions (Q1-Q10) gathered BEV users' demographic and driving/vehicle-related information. Specifically, based on preliminary research (Wang et al., 2023), except for basic demographic information (Age, gender, and Income), we selected the trust to BEVs' RES (BEV Trust), driving experience (Driving Distance), and driving regional difference (Infrastructure and Temperature) as the potential factors to charging behavior and range anxiety. Moreover, the individual differences in personality traits were included following previous research on range anxiety (Rauh et al., 2015).

- Q11 was designed to measure BEV users' *Comfort Time*, which is a construct that measures one's sensitivity to the waiting time before charging. The *Comfort Time* can reflect one's time-related range anxiety level and might be related to one's charging decision.

- Q12 aimed to measure drivers' comfort range, which is related to distance-related anxiety (Yuan et al., 2018). Three comfort ranges were assessed for different lengths of trips, given that the relationship between comfort range and trip length may not be linear.
- Q13 presents scenarios in which participants needed to make charging decisions. Following Wang et al., (2023), the scenarios were defined by three factors (i.e., *Waiting Time*, *State-of-Charge (SoC)*, and *Rest Trip*). Specifically, the *Waiting Time* is the time drivers need to wait for charging at the upcoming charging station. The *SoC* is the current rest battery level (%) of BEV, and the *Rest Trip* is the remaining distance to the destination. It is worth noting that, the *Rest Trip is* presented to drivers after multiplying the *SoC*, which aims to control the perceived distance-related anxiety that drivers are experiencing (i.e., a higher *Rest Trip* means the remaining distance to the destination will consume more energy). Thus, the trade-offs between distance-based and time-based range anxiety can be evaluated. It should also be noted that all three scenario-related factors were allowed to vary within specific ranges (see Table 2). Thus, the scenarios each participant encountered may not be the same, so that we can observe the continuous influence of a factor on the charging decision. In total, we constructed 60 scenarios for each participant (5 *Waiting Time* ranges * 3 *SoC* ranges * 4 *Rest Trip* ranges).

In the third column of Table 2, we provide the distribution of each variable and how each level of the discrete variable was labeled. To check the credibility of the collected data, we conducted a reliability and validity assessment of the responses to standard questionnaires (i.e., Q4 and Q9) using Cronbach's alpha, Kaiser-Meyer-Olkin measure (KMO), and Bartlett's test of sphericity (Paltun & Bölükbaş, 2021). All these metrics reached a satisfactory level (see Table 2). Given that the study focused on BEV users in mainland China, a Chinese version of the questionnaire was utilized. All questions were translated into Chinese if no standard translations were available; otherwise, the standard Chinese version of the questionnaire was used (for Q4).

**Table 2.** Questions, Extracted Variables, Distribution of the Data

| Questions | Variables | Distribution ~ label in ML model |
|---|---|---|
| Q1: [FI] Date of birth. | *Age* | Mean: 26.4 years old (SD: 4.7, min: 18, max: 40) |
| Q2: [SC] Gender at birth. | *Gender* | • Female (n=55, 24.0%) ~ 1<br>• Male (n=175, 76.0%) ~ 0 |
| Q3: [SC] Please describe your annual family income level. | *Income* | • ≥ 40K (n=128, 55.4%) ~ 2<br>• ≥ 14K & < 40K (n=44, 19.4%) ~ 1<br>• < 14K (n=58, 25.2%) ~ 0 |
| Q4: [LS] The Ten Item Personality Questionnaire (TIPI) (Franke et al., 2015)<br>- Cronbach α = 0.839<br>- KMO = 0.831<br>- *p* value of Bartlett's Sphericity test <.0001 | *Extraversion*<br>*Agreeableness*<br>*Conscientiousness*<br>*Emotional Stability*<br>*Openness to Experiences* | Mean: 4.4 (SD: 0.9, min: 2.0, max: 7.0)<br>Mean: 4.9 (SD: 1.1, min: 2.0, max: 7.0)<br>Mean: 5.1 (SD: 1.2, min: 2.0, max: 7.0)<br>Mean: 4.9 (SD: 1.2, min: 2.0, max: 7.0)<br>Mean: 4.7 (SD: 1.1, min: 2.0, max: 7.0) |
| Q5: [FI] Please indicate the province you drive the most.<br>- Further categorized into three levels based on EV infrastructure development (Cheng Zheng, 2021).<br>- Further categorized into three levels based on annual average temperature (*Surface Climate Diagram of China*, n.d.). | *Infrastructure* | • Well developed (n=105, 45.5%) ~ 2<br>• Average (n=97, 42.1%) ~ 1<br>• Less developed (n=28, 12.4%) ~ 0 |
| | *Temperature* | • South (n=134, 58.3%) ~ 2<br>• Central (n=56, 24.5%) ~ 1<br>• North (n=40, 17.2%) ~ 0 |
| Q6: [SC] Please indicate the driving distance (km) in the recent year of the BEV you drive the most. | *Driving Distance* | • ≥ 30k (n=52, 22.7%) ~ 2<br>• ≥ 10k & < 30k (n=145, 62.8%) ~ 1<br>• < 10k (n=33, 14.5%) ~ 0 |
| Q8: [SC] Please indicate the mileage computation mode of the BEV you drive the most. | *Range Estimation Cycle* | • I don't know (n=43, 18.7%) ~ 4<br>• WLTC (n=56, 24.0%) ~ 3<br>• EPA (n=16, 7.0%) ~ 2<br>• NEDC (n=68, 29.7%) ~ 1<br>• CLTC (n=47, 20.6%) ~ 0 |
| Q9: [LS] Trustworthiness scale (FIFT) (Franke et al., 2015) regarding users' trust in RES of the BEV they drive the most.<br>- 1 ("not at all") to 6 ("extremely")<br>- Cronbach α = 0.744<br>- KMO = 0.796<br>- *p* value of Bartlett's Sphericity test <.0001 | *BEV Trust* | Mean: 4.0 (SD: 0.8, min: 1.6, max: 6.0) |
| Q10: [SC] What is the maximum DISPLAY mileage (km) of your BEVs when fully charged? | *Display Mileage* | • Over 550 (n=18, 7.9%) ~ 3<br>• [350, 450] (n=54, 23.6%) ~ 2<br>• [350, 450] (n=121, 52.6%) ~ 1<br>• [250, 350] (n= 37, 15.9%) ~ 0 |
| Q11: [FI] For a highway trip that is beyond the real mileage of a BEV (i.e., you will need to recharge once in the middle of the trip). If the waiting time before charging is *t* minutes, would you choose a BEV or a fuel car?<br>- 120 ≥ *t* ≥ 0 minutes | *Comfort Time* | mean: 43.8 minutes (SD: 33.6, min: 0, max: 100) |
| Q12: [SSC] If the trip is *m* km and there are no charging stations along the way, what is your minimum comfortable percent of display mileage before the trip starts?<br>- *m* = [25%, 50%, 75%] * *Display Mileage* | *Comfort Range 25%* | mean: 47.3% (SD: 16.4, min: 25.0, max: 100) |
| | *Comfort Range 50%* | mean: 70.7% (SD: 12.2, min: 52.0, max: 100) |
| | *Comfort Range 75%* | mean: 88.6% (SD: 7.0, min: 75.9, max: 100) |
| Q13: [SSC] You are driving on the highway. When approaching an upcoming rest area, the navigation informs you that the waiting time before charging at the area is *t* minutes, the remaining battery range of the BEV is *r* km and you are *d* km away from destination (where you have plenty of time to recharge), would you choose to charge at this area or charge at the destination?<br>- *t* = [0-20, 20-40, 40-60, 60-80, 80-100]<br>- *r* = [20-40%, 40-60%, 60-80%] * Display Mileage<br>- *d* = [15-35%, 35-55%, 55-75%, 75-95%] * *r* | *Waiting Time (t)* | [0-20, 20-40, 40-60, 60-80, 80-100] min |
| | *SoC (r)* | [20-40%, 40-60%, 60-80%] |
| | *Rest Trip (d)* | [15-35%, 35-55%, 55-75%, 75-95%] |
| | *Charge Decision* | • Charge at the destination (55.7%) ~ 1<br>• Charge at the upcoming rest area (44.3%) ~ 0 |

*Note: Abbreviations of question types are as follow: FI: Fill-in-text; SC: Single-choice; MC: Multiple-choice; LS: Likert scale, SSC: Scenario Single-choice; TF: True or false. SD standards for standard deviation.'~' follows the label of each level in ML model construction.*

*Participants*

All participants were recruited online. Responses from commercial BEV drivers (i.e., ride-hailing, taxi, cargo services) were removed, as they might have developed different charging strategies compared to those who drive BEVs for personal purposes. A total of 287 participants completed the questionnaire. Then, all responses were screened based on two quality-checking questions (57 samples were excluded). Ultimately, 230 valid responses were used for analysis (175 males and 55 females), leading to 13,800 charging decision samples (230 participants*60 scenarios/participant). According to Gasgoo (2019), our sampled gender ratio is close to the real-world BEV user group portrait (i.e., the ratio of male to female is close to 7:3). Participants were compensated with 5 RMB for their completion of the 15-minute questionnaire. The research protocol was approved by the Human and Artefacts Research Ethics Committee at the [**Place holder for double-blind review**] (protocol number: [**Place holder for double-blind review**]).

*ML models and Comparison Experiment*

The goal of ML models is to predict the drivers' charging decisions in different scenarios. In this study, we defined the charging behavior with two choices (i.e., charge at the destination or charge at the upcoming rest area). Therefore, it is a classic binary classification task. We adopted seven common supervised learning ML models to predict participants' charging decisions (i.e., charge at the destination or at the upcoming rest area) in the scenarios assessed in Q13:

**LightGBM** was first proposed by Ke et al., (2017). It is a gradient-boosting framework that uses a tree-based learning algorithm. Attributed to the amalgamation of pioneering techniques (i.e., Gradient-based Single-Side Sampling and Exclusive Feature Bundling), LightGBM has demonstrated superior performance in classification tasks. It can naturally handle missing values, efficiently encode categorical features, and prevent overfitting with with parameters such as leaf depth limits.

**Logistic Regression (LR)** (Hosmer Jr et al., 2013) is a supervised learning technique that employs a generalized linear regression model to estimate the probability of a sample belonging to a specific class. The process of implementing LR involves identifying a prediction function, generating a loss function, and determining regression parameters that minimize the loss function. The primary objective of LR is to establish the relationship between the dependent and explanatory variables.

**AdaBoost (Adaptive Boosting)** (Freund & Schapire, 1997) is an ensemble learning algorithm that combines multiple weak classifiers to create a strong classifier. It works by iteratively training weak classifiers on different subsets of the data, with each subsequent classifier giving more weight to the misclassified samples from the previous classifier.

**Decision Tree (DT)** (Safavian & Landgrebe, 1991) is a supervised learning algorithm that uses a tree-like model to make predictions. Each internal node in the tree represents a decision based on a specific feature, while the leaf nodes represent the outcomes or predictions. DT is easy to interpret and can handle both numerical and categorical data. They automatically select the most important features for splitting and can handle complex nonlinear relationships.

**Random Forest (RF)** (Breiman, 2001) is an ensemble learning technique that involves the construction of multiple DT classifiers and the aggregation of their results (Sideris et al., 2019). Then RF makes predictions based on the aggregated results. RF has been widely adopted in previous research, given its high adaptability and ease of use for solving both regression and classification problems.

**XGBoost** (Chen and Guestrin 2016) is a tree-based ensemble machine learning model. XGBoost operates on the "boosting" principle, which leverages additive training methods to integrate the predictions of weak learners and create a robust learner (Ullah et al., 2022). The XGboost has been widely implemented across various fields and has demonstrated its effectiveness as an ensemble model.

**Multi-Layered Perception (MLP)** (Jordan & others, 1995), also known as neural networks, is a deep learning algorithm that is inspired by the structure and function of the human brain. It consists of multiple layers of interconnected nodes (or neurons) that process and transforms input data to produce outputs. MLP can be used for a wide range of tasks, including image and speech recognition, natural language processing, and predictive analytics.

We first split the dataset into a training dataset and test dataset with a ratio of 4:1. Then, to fairly compare the model performance and select the best one, all models were tuned using the 5-fold cross-validation. In the 5-fold cross-validation, the whole dataset set was randomly divided into five equal-size subsets, and four subsets were used to train the model, while the remaining subset was retained for validation. This was repeated five times, with each subset being used as the validation dataset once. The five outcomes on the validation dataset were then averaged to evaluate the model performance.

Next, to select the most informative variables for further analysis, the variable selection was performed based on Recursive Feature Elimination (RFE). RFE is a wrapper feature selection method that eliminates features recursively (Pedregosa et al., 2011). In general, it eliminates features with the least information one by one greedily until it finds the optimal feature subset space. Specifically, in this study, we obtained the best-fitted model after the performance comparison over seven ML methods. Then, we initialized the RFE component based on the best-fitted model for all variables. Model construction, training, hyper-parameter tuning, and variable selection were conducted using the Scikit-learn library (Pedregosa et al., 2011). Lastly, RFE was based on the model performance on the test dataset.

*SHAP Analysis*

Shapley Additive exPlanations (SHAP) is a model interpretability method proposed by Lundberg and Lee (2017) that provides localized interpretations for individual predictions out of ML models. Using the Shapley value from cooperative game theory, SHAP allocates each feature to the portion of a model prediction that is attributable to that feature. The Shapley value, $\varphi_i$, represents the fair, unique solution for distributing gains among players in a cooperative game. Assuming the model has a feature set N, the Shapley value of i-th feature in N is:

$$\varphi_i = \sum_{S \subseteq N\{i\}} \frac{|S|!\,(|N|-|S|-1)!}{N} [f(S \cup \{i\}) - f(S)] \quad (1)$$

Where $f(*)$ represents the predicted values based on given features, N{i} means the subset of N after removing i-th feature, and S is the given feature set. Each feature is considered a "contributor" to the model's predictions. In general, SHAP values provide a local, model-agnostic interpretation of each prediction made by the model. SHAP values can be positive or negative for each feature. In our study, for the ML model that reached the best performance in the comparison experiment, we re-trained the model on the whole dataset and then conducted the SHAP analysis.

*Bayesian network-regression modeling*

The Bayesian network (BN) is a graphical probabilistic model that employs Bayes' theorem to express conditional dependencies between variables (Hosmer Jr et al., 2013). The model is represented by a directed acyclic graph (DAG), where each variable is represented as a node, and the relationships between nodes are defined as directed edges. The confidential dependency in the edges can be observed from the dataset or pre-set by prior knowledge (Heckerman, 2008). There are three approaches for constructing the BN structure (Sun & Erath, 2015): data-based, prior-knowledge-based, and hybrid. While the data-based approach can provide informative structures and good prediction performance, it may be limited by the quality and quantity of the data (Khakzad et al., 2011). At the same time, the prior-knowledge-based approach may not be able to accurately identify the dependency structure. Therefore, the hybrid approach was chosen in our study, which mixed data-based and prior-knowledge-based approaches to build the BN structure.

Specifically, in the BN modeling, we first group factors that cannot be dynamically changed by others and can be measured objectively into a layer (i.e., *Age, Display Mileage,* and *Driving Distance*) and factors that can hardly be measured into another layer (i.e., *Trust and* personality-related factors). Then, a fully connected network was initiated, with all factors outside these two layers linked with each other; all factors outside these two layers were also linked to all factors in these two layers. Next, the initial network was pruned based on the data-driven automated constraint conditional dependency searching (Schulte et al., 2009). Only edges (i.e., links connecting two factors) with significant ($p<.05$) conditional dependences in the chi-squared tests were retained in the BN. The "*pgmpy*" package (Ankan & Panda, 2015) in Python 3.8 was used to build the BN structure.

Finally, to quantify the relationships among influential variables, regression analyses were conducted for all hierarchical sub-structures in BN. Mixed linear regression models (using Proc MIXED procedure) were implemented in "SAS OnDemand for Academics". Specifically, for sub-structures in BN, we built regression models with the node itself as the dependent variable and all its parental nodes as independent variables. To avoid the multicollinearity problem, we adopted backward stepwise selection procedures based on model fitting criteria and Variance Inflation Factor (e.g., Infrastructure was kept, but Temperature was abandoned in the model of Knowledge). The Tukey-Kramer post-hoc tests (KRAMERß, 1956) were conducted for all significant variables ($p < .05$) in each sub-structure.

## Results

### *Comparisons of ML models*

In this study, we compared the performance of seven ML algorithms in predicting BEV charging decisions. To comprehensively evaluate the machine learning models, we employed various classification performance metrics, including accuracy, F1 score, and area under the receiver operating characteristic curve (AUC) score. Specifically, accuracy measures the ratio of correctly classified samples to the total number of samples. The F1 score calculates the weighted average of precision and recall, and it performs better in evaluating the model performance compared to accuracy, especially when the dataset has an imbalanced distribution of classes. As for AUC, a higher score indicates better classification performance. The detailed results are reported in Table 3 and Figure 2.

**Table 3.** Models' Performance Comparison

|  | LightGBM | LR | DT | AdaBoost | RF | XGBoost | MLP |
|---|---|---|---|---|---|---|---|
| Accuracy (%) | **77.48** | 63.57 | 70.52 | 68.89 | 75.35 | 76.22 | 68.65 |
| F1 score (%) | **79.66** | 69.67 | 73.58 | 73.03 | 77.95 | 78.64 | 73.81 |
| AUC score | **0.857** | 0.691 | 0.698 | 0.749 | 0.836 | 0.847 | 0.775 |

*Notes: The highest scores are bolded in **black**.*

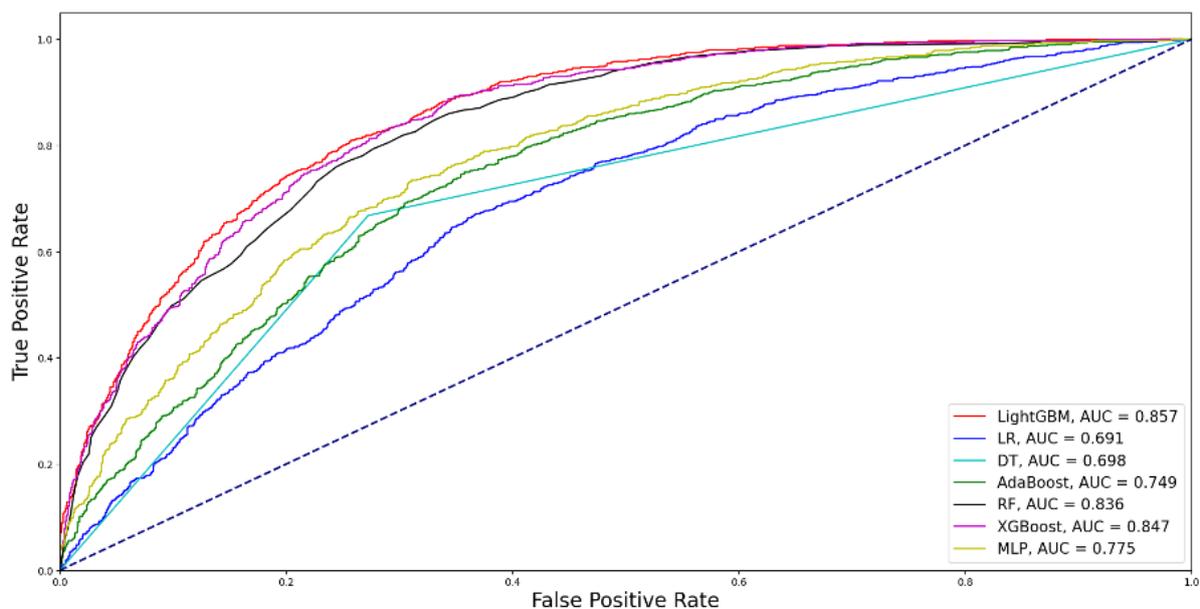

**Figure 2.** Receiver operating characteristic (ROC) curves of ML methods

The results in Table 3 and Fig. 2 show that LightGBM achieved the highest accuracy, F1 score and AUC among all models, indicating that the LightGBM outperformed all other candidate ML models explored in this study. DT outperformed AdaBoost in terms of accuracy and F1 core, though DT yielded lower AUC scores, indicating that models with higher complexity do not always outperform simpler models. Besides, we observed that tree-based models (i.e., LightGBM, DT, Adaboost, RF, and XGBoost) always achieve higher accuracy than others. It seems that the tree-based model has better fitting capacity in tableau data. Overall, based on this result, we chose LightGBM as the final ML model for further analysis. The final hyper-parameters of LightGBM are listed in the following table:

Table 4. Hyper-parameters of the final LightGBM model

| Parameter | Description | Values |
| --- | --- | --- |
| n_estimators | The number of boosting iterations. | 200 |
| max_depth | The maximum number of splits for base learners (-1 means no limit). | -1 |
| subsample | The fraction of observations which randomly selected for training. | 0.8 |
| subsample_freq | Frequency for bagging. | 100 |
| learning_rate | The model learning rate. | 0.1 |
| min_split_gain | The minimum loss reduction required to perform a split. | 0 |
| reg_lambda | L2 regularization term. | 0 |
| reg_alpha | L1 regularization term. | 0.8 |

**Results of SHAP Analysis**

*Factor importance analysis*

To gain a better understanding of influential factors of BEV charging behaviors, a SHAP analysis was conducted for the ML model with the best performance. Before the SHAP analysis, a feature selection was conducted based on RFE. Then, We evaluated the relative importance of the remaining influential factors in the LightGBM model. Fig. 3a visualizes the average absolute impact of individual feature (or factor) on the model output magnitude: higher relative importance indicates that a factor exerts a larger impact on the charging decision.

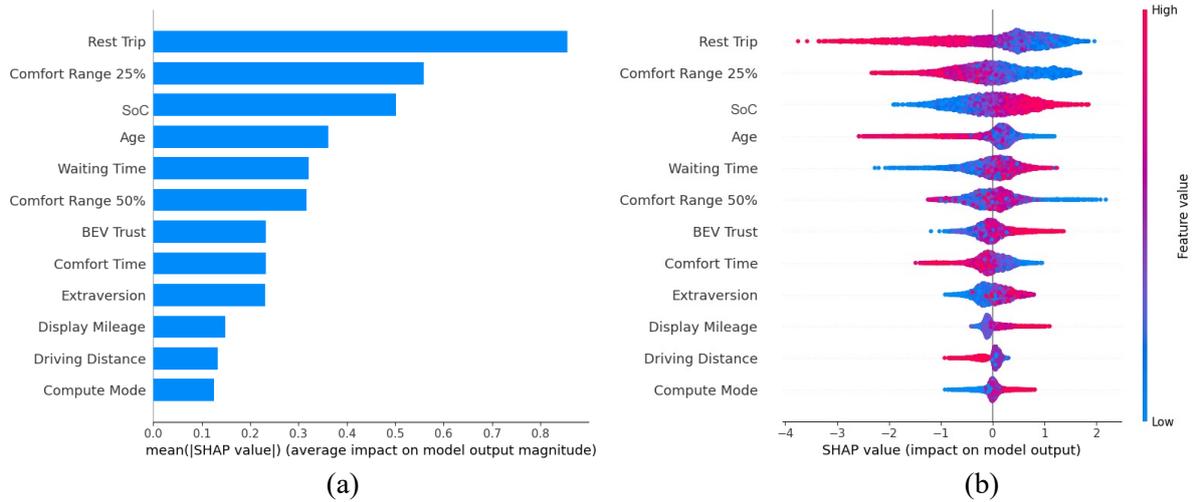

**Figure 3.** Summary plot of SHAP values of influential factors

Fig. 3b further presents a summary plot of SHAP values for the two charging decisions (i.e., charge at the destination or charge at the upcoming rest area). The factors are ranked in descending order of importance on the y-axis, while the SHAP values indicating the effect of each factor on the charging behaviors are shown on the x-axis. In SHAP analysis, a factor associated with a higher SHAP value is more likely to be associated with the decision of charging at the destination and vice versa. A SHAP value of 0 means the factor has no impact on users' charging decisions. The color bar depicts the value of each factor, with blue indicating a lower value and red indicating a higher value. For example, the lower *Rest Trip*, *Comfort Range25%*, and *Age* were associated with a higher likelihood of charging at the destination; while lower *Waiting Time* and *SoC* were associated with a higher likelihood of charging at the upcoming rest area.

*Main effects and interaction effects*

To better reveal the relationships between the factor values and the charging decisions, SHAP dependence plots were provided in Figure 4 and Figure 5 for selected main and interaction effects. The SHAP dependence plots of other interaction effects are provided in Appendix. In the SHAP dependence plots, the influence of a factor on the charging decision is depicted as the vertical dispersion of SHAP values. Particularly, following Islam & Abdel-Aty (2023), to better illustrate the trends of SHAP values for interpretation, we fitted the SHAP values of each continuous factor with first-order polynomial functions through the *polyfit* function in *Numpy* package, which was visualized as red lines in the figures.

By examining the corresponding factor values when the fitted lines crossed the horizontal line of 0 SHAP value, we can determine when users' decisions would change, i.e., the decision boundary.

In Figure 4, the main effects of influential factors identified in the feature selection are illustrated. For three scenario-related factors, firstly, a larger *Rest Trip* was found to be associated with a higher likelihood of charging at the upcoming rest area (Figure 4a). *Rest Trip* was the remaining distance to the destination of the travel. The decision boundary of *Rest Trip* was about 55%, indicating drivers tend to charge nearby when the remaining driving distance is lower than the 55% maximum range of their BEVs. At the same time, we found that an increase in *SoC* (i.e., the rest battery level in that scenario) was associated with a higher likelihood of charging at the destination; the decision boundary was around 50% level of the battery (Figure 4c). Finally, it was found that as *Waiting Time* (i.e., the waiting time before charging) increases, SHAP increases and stabilizes at a positive value (i.e., more likely to charge at the destination). Specifically, when *Waiting Time* was below 45 minutes, the SHAP values were mostly negative, indicating that the users preferred to charge at the upcoming rest area (Figure 4f).

The influence of the social-psychological factors is mixed. The trends of *Age* (Figure 4e), *Comfort Range 25%* (Figure 4b), *Comfort Time* (Figure 4i), and *Driving Distance* (Figure 4l) are similar. In general, older drivers, those who had a higher comfort range at the battery level of 25%, those who reported longer comfort time (i.e., one's sensitivity to the waiting time before charging and the shorter, the higher time-related anxiety one has) or higher *Comfort Range 25%* (i.e., one's comfortable battery range when the trip takes 25% of displayed battery range and the shorter, the lower distance-related range anxiety one is), and those who had longer BEV driving distance last year were more likely to charge at the upcoming rest area. At the same time, in general, higher *BEV Trust* and *Extraversion* were found to be associated with an increased preference to charge at the destination. Further, as expected, drivers owning a BEV with lower *Display Mileage* were more likely to charge at the upcoming rest area. The influence of the *Range Estimation Cycle* (REC) is interesting. According to some comparison experiments (Bilal Akgunduz, 2021), the CLTC is the least accurate, followed by NEDC, then EPA, and WLTC. It seems that, in general, the more accurate the display mode is, the more likely drivers prefer to charge at the destination. Drivers who did not know their display mode, however, were the most likely to charge at the destination. Finally, *Comfort Range 50%* had no clear linear influence on users' charging decisions.

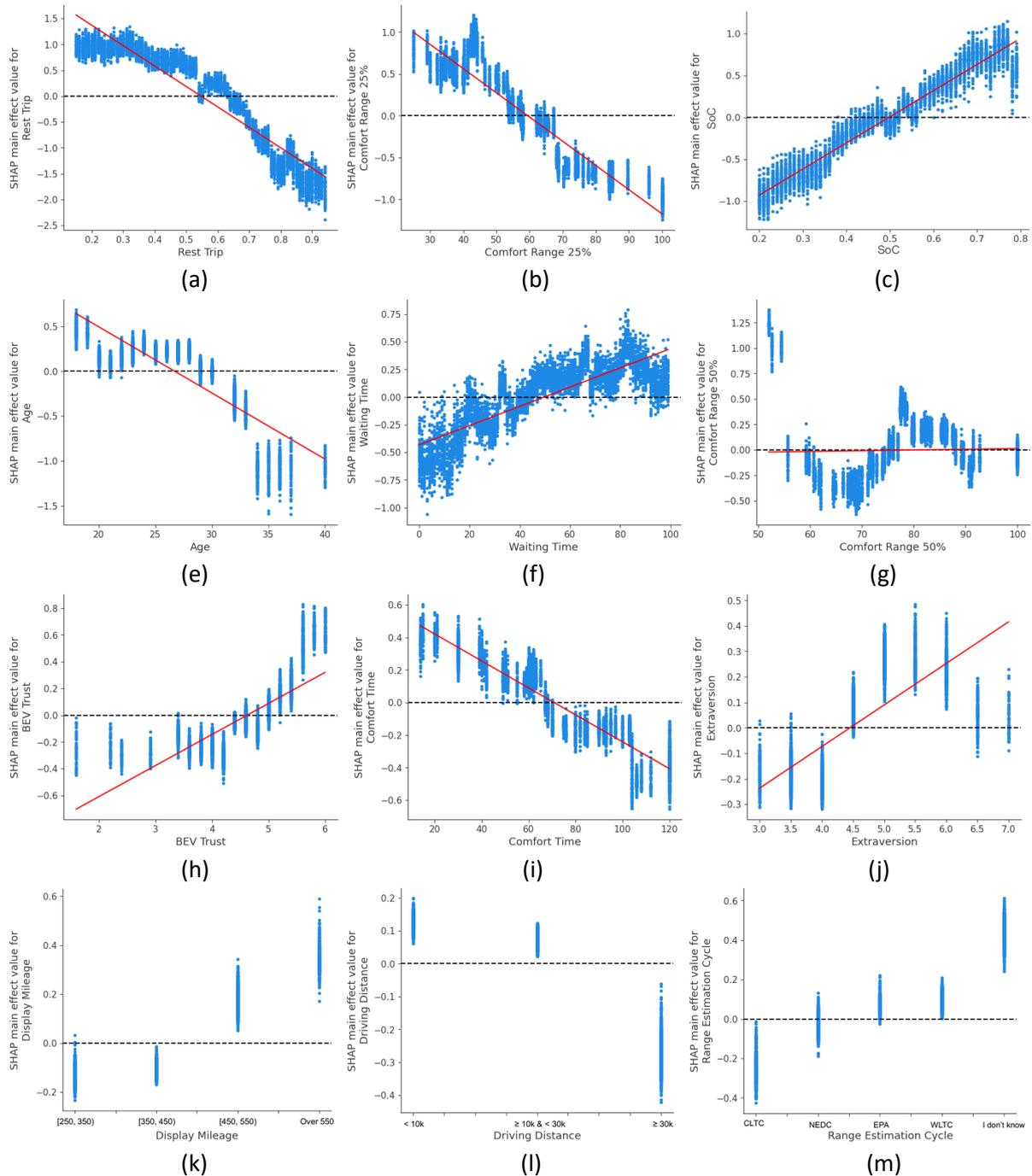

**Figure 4.** SHAP main effects plots of influential variables

To better understand the potential joint effects of the identified influential scenario-related factors (i.e., *SoC*, *Rest Trip*, and *Waiting Time*) and range-anxiety-related factors (*Comfort Range 25%*, *Comfort Range 50%*, and *Comfort Time*). We conducted interaction effects analyses. In Figure 5. the x-axis represents the value of one factor (main factor), while the value of another factor (secondary factor) is illustrated using different colors. In the plot, the distribution of the SHAP value given the main factor can be obtained by observing the distribution of the dots at a specific horizontal location. Still, the larger the SHAP value, the more likely the users choose to charge at the destination. It is interesting to find that some factors can affect users' decisions regardless of the level of other factors. For example, as shown in Figure 5a, i.e., the distribution of the SHAP values always crosses 0, given any value of waiting time, though the exact values may vary. In other words, users' charging decisions always changed with the *SoC*. While for some other factors, it may only affect users' charging decisions when the values of other factors are within a specific range. For example, in Figure 5i, when the *Rest Trip* is large, users would always prefer to charge at the upcoming rest area, regardless of uses' *Comfort Time*.

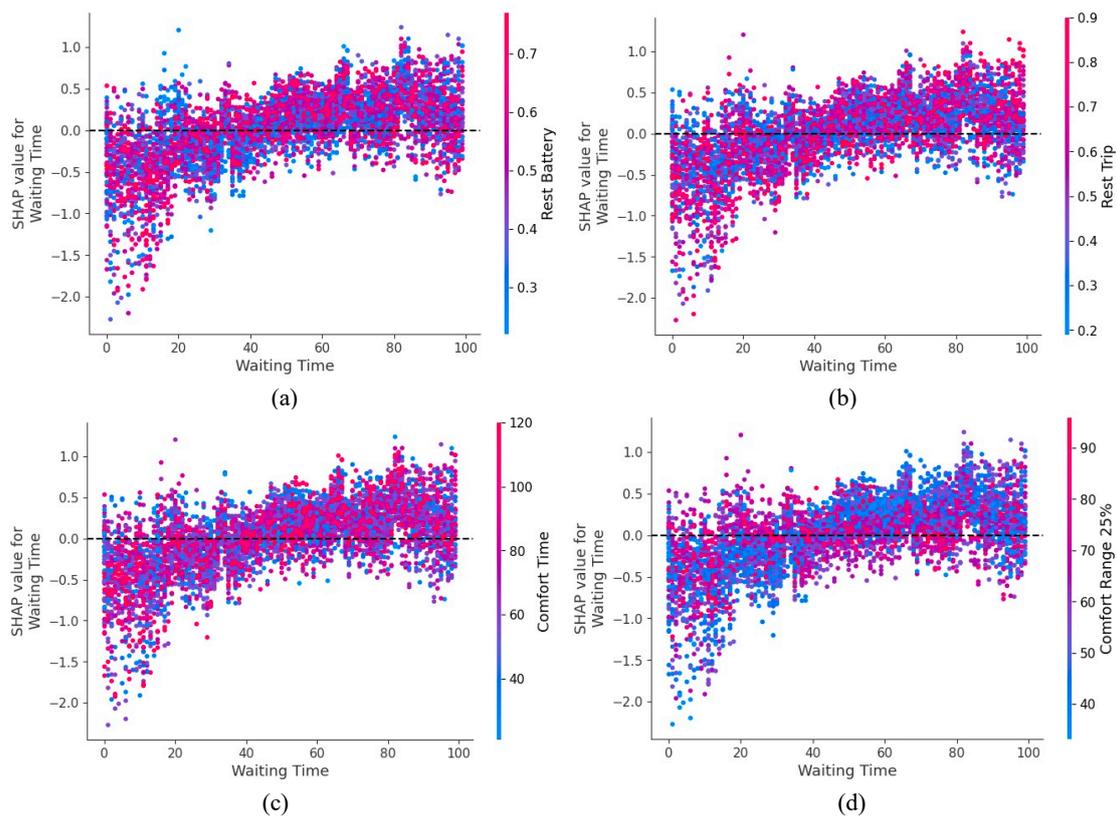

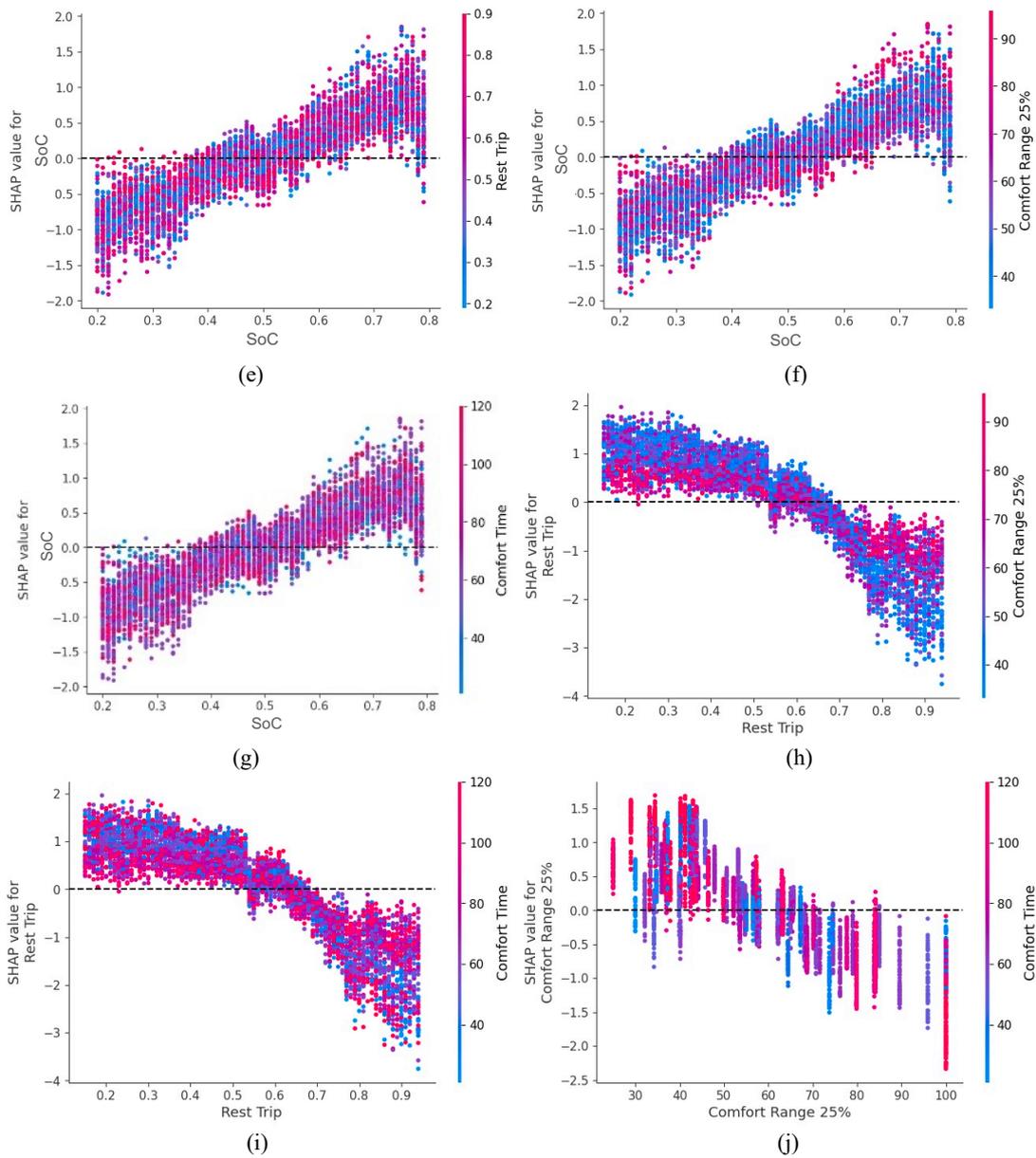

**Figure 5.** SHAP interaction effects plots: a) Interaction between Waiting Time and SoC; b) Interaction between Waiting Time and Rest Trip; c) Interaction between Waiting Time and Comfort Range 25%; d) Interaction between Waiting Time and Comfort Time; e) Interaction between SoC and Rest Trip; f) Interaction between SoC and Comfort Range 25%; g) Interaction between SoC and Comfort Time; h) Interaction between Rest Trip and Comfort Range 25%; i) Interaction between Rest Trip and Comfort Time; j) Interaction between Comfort Range 25% and Comfort Time.

## *Results of BN-regression Modeling*

To further understand inter-correlations among the influential factors, a BN-regression analysis was conducted. The final DAG is presented in Fig. 6. A three-layered structure was observed. The blue box consists of range anxiety-related factors with internal dependencies. The factors in the green box are the demographic factor (*Age*), driving experience (*Driving Distance*), and BEV-related factors (*Display Mileage*) that can be measured objectively. The orange box contains self-reported personality and trust in BEVs' RES, which are psychological factors that are relatively stable but can hardly be measured objectively. Note that BN cannot inform causal relationships; thus the arrows in Figure 6 were determined by prior knowledge. Specifically, range anxiety-related factors could be influenced by all social-psychological factors (the ones in green and orange boxes), but range anxiety factors cannot change social-psychological factors.

Then, the influence of each arrow was assessed through statistical regression models. Regression models were built for range-anxiety-related factors, i.e., with *Comfort Range 25%*, *Comfort Range 25%*, and *Comfort Time* as the dependent variable for each model, respectively. Those arrows with significant relationships (*p*-value < .05) in the regression models were highlighted in red in Figure 5, with detailed statistical results shown in Table 5. The influence of the continuous independent variables on the range-anxiety related factors can be found in Table 5, while for the categorical independent variables, their association with the range-anxiety-related factors is provided as post-hoc contrasts in Table 6.

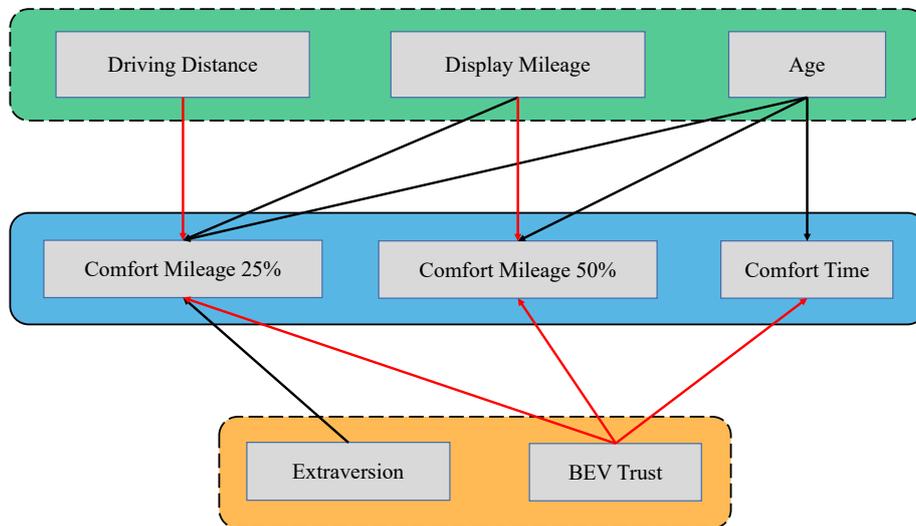

**Figure 6.** The Final DAG. Note that, all three factors in the blue box are fully connected, while no connections were observed within the green and orange boxes.

**Table 5.** Summary of statistical results

| Dependent Variable (DV) | Independent Variable (IV) | F-value | Estimate (95% CI) | *p*-value |
|---|---|---|---|---|
| Comfort Range 25% | Driving Distance | F(2, 192) = 4.37 | - | .014 * |
| | Comfort Time | F(1, 192) = 13.54 | 0.13 [0.06, 0.20] | .0003 * |
| | BEV Trust | F(1, 192) = 4.38 | 3.04 [0.18, 5.90] | .04 * |
| | Age | F(1, 192) = 0.13 | 0.09 [-0.38, 0.56] | .7 |
| | Display Mileage | F(3, 192) = 2.27 | - | .08 |

| | Extraversion | F(1, 192) = 0.05 | -0.27 [-2.69, 2.14] | .8 |
| --- | --- | --- | --- | --- |
| Comfort Range 50% | BEV Trust | F(1, 195) = 5.30 | 2.37 [0.34, 4.39] | .02 * |
| | Age | F(1, 195) = 2.80 | 0.28 [-0.05, 0.61] | .095 |
| | Comfort Time | F(1, 195) = 17.30 | 0.10 [0.05, 0.15] | <.0001 * |
| | Display Mileage | F(3, 195) = 7.09 | - | .0002 * |
| Comfort_Time | BEV Trust | F(1, 198) = 30.83 | 14.68 [9.47, 19.90] | <.0001 * |
| | Age | F(1, 198) = 1.19 | -0.50 [-1.40, 0.40] | .3 |
| | Comfort Range 25% | F(1, 198) = 17.23 | 0.54 [0.29, 0.80] | <.0001 * |

*Note: '-' means the post-hoc contrasts are provided in Table 6.*

**Table 6.** Significant Post-hoc results for discrete independent variables

| DV | IV | IV Level | IV Level compared to | Δ (95% CI) | t value | *p*-value |
| --- | --- | --- | --- | --- | --- | --- |
| Comfort Range 25% | Driving Distance | < 1W | ⩾ 1W & < 3W | -2.42 [-10.10, 5.26] | t(192) = -0.74 | .7 |
| | | | ⩾ 3W | -9.63 [-18.64, -0.63] | t(192) = -2.53 | .03 * |
| | | ⩾ 1W & < 3W | ⩾ 3W | -7.21 [-13.60, -0.82] | t(192) = -2.66 | .02 * |
| Comfort Range 50% | Display Mileage | [250, 350) | [350, 450) | -4.14 [-9.93, 1.65] | t(195) = -1.85 | .07 |
| | | | [450, 550) | -8.69 [-15.16, -2.21] | t(195) = -3.48 | .004 * |
| | | | Over 550 | -13.0 [-21.62, -4.29] | t(195) = -3.88 | .003* |
| | | [350, 450) | [450, 550) | -4.55 [-9.46, 0.37] | t(195) = -2.40 | .08 |
| | | | Over 550 | -8.81 [-16.37, -1.26] | t(195) = -3.02 | .02 * |
| | | [450, 550) | Over 550 | -4.27 [-12.39, 3.85] | t(195) = -1.36 | .2 |

*Note: Δ = IV Level - IV Level compared to: when it is positive, it means IV Level > IV Level compared to and vice versa.*

# Discussion

In this study, we evaluated drivers' charging decisions in scenarios where time-related anxiety and distance-related anxiety co-exist. Further, we analyzed the factors influencing users' decisions using mixed approach combining ML, SHAP, BN and regression analysis and revealed the association among the influential factors of BEV users' charging decisions.

First of all, for the first time, our study reveals that both distance-related factors and time-related factors can influence users' charging decisions. Specifically, with the increase in the *Rest Trip*, drivers tended to charge earlier, even though the displayed mileage was longer than the rest trip, indicating that the distance-related range anxiety alone may not fully reveal users' psychological states. At the same time, with the increase in the waiting time at the upcoming charging area, drivers tended to delay charging their BEVs – the influence of the waiting time saturated at around 45 minutes in general. Previous research mostly considered only distance-related factors when optimizing the charging network or recommending charging strategies (Yan & Tang, 2023), our study suggests that ignoring the time-related factors may have over-simplified BEV users' decision process. Additionally, using a more advanced framework (i.e., ML and BN mixed approach), we were able to reveal the underlying relationship of the factors leading to a charging decision and explore how individual differences (materialized as *Extraversion, Driving Distance*, and *Age*) may moderate one's decision. Thus, this research can provide insights on not just model one's decision, but also on how to affect one's decision with one's individual differences considered.

At the same time, our study reveals the role of range anxiety on users' charging decisions. In (Wang et al., 2023), only interaction effects of *Comfort Time* were identified, but there were no significant linear relationships between *Comfort Time* and charging decisions found because of the limitations in linear regression models. By contrast, through our novel analysis framework (refer to Fig. 1), we quantified two types of range anxiety using two variables: *Comfort Range* for distance-related anxiety and *Comfort Time* for time-related anxiety - both played significant roles in predicting users' charging decisions. As expected, the drivers who could tolerate longer *Waiting Time* before charging tended to charge at the upcoming charging area, even if they had to wait. Regarding the influence of the comfort ranges, we found that, compared to users' self-reported comfort range when the battery level was high (50% or higher), users' comfort range when the battery level was low (25%) was more strongly associated with users' charging decisions. The seemingly non-linear relationships between *Comfort Range 50%* and charging decisions might indirectly affect drivers' charging behavior when the rest of the battery level is more adequate. This finding was missing in (Wang et al., 2023), and further mining on possible intermediate influential factors in that circumstance is expected. Previous research that evaluated drivers' distance-related range anxiety covered the SoC level from 5% to 45% (Yuan et al., 2018); our research suggests that users' comfort range at higher battery levels affected little on users' charging decisions. Thus, when training models to predict users' charging decisions, we may put more weight on users' decisions when the remaining distance is low. Further, from the perspective of rest area management, with the knowledge of the effects and leading factors of range anxiety, the charging recommendation strategy can be designed based on the driver's personalized psychological preferences and charging decision boundaries, which may alleviate driver's distance-related and time-related range anxiety and at the same time, help avoid long queuing and waiting times at some specific charging stations.

In addition to the range anxiety-related factors, users' characteristics may also partially explain users' decisions. Previous research pointed out that users' trust in the range estimation systems (RESs) of BEV can affect users' range anxiety (Rauh et al., 2015). Our research further validates that users' trust in RES is associated with users' decisions. What is more interesting, through the BN-regression mixed approach, our study reveals that trust may affect users' charging decisions by affecting users' time-related and distance-related anxieties. With the increase of trust in RES, users would have higher levels of *Comfort Range 25%*, *Comfort Range 50%*, and *Comfort Time*. The positive association between range anxiety (i.e., *Comfort Range 25%*, *Comfort Range 50%*) and trust is straightforward – those who trusted the RES more might perceive the RES as more reliable (Merritt et al., 2013) and thus were more confident that they could reach the destination. This trend has also been further confirmed by how the *Range Estimation Cycle* was associated with users' charging decisions – in general, the more accurate the estimation, the more likely drivers choose to charge at the destination. The positive association between trust in RES and *Comfort Time*, however, might be explained by the existence of potential covariants that are not explored in our study. Future research may need to investigate more demographic factors to better reveal these relationships observed in our study. In addition to Trust, we found that the *Age* and *Extraversion* of drivers can also affect users' charging decisions. It seems that older drivers and those who were less extroverted placed more weight on distance-related factors in terms of making charging decisions – they preferred to charge whenever possible, even if they had to wait. The effects of age and personality on range anxiety have been observed in previous studies (Franke et al., 2012; Yuan et al., 2018), but we further revealed their roles in moderating charging decisions.

It should be noted that the influence of some factors might be easily moderated by other factors, while the influence of some other factors may dominate users' choices. For example, it was found that the influence of *Waiting Time* is softer compared to other scenario-related factors (i.e., *SoC* and *Rest Trip*). Regardless of how long the *Waiting Time* is, users' choice could be affected by other factors (e.g., *Rest Trip* and *SoC*); while when the *SoC* level is low or when the *Rest Trip* is high, drivers would always prefer to charge at the upcoming rest area, no matter how long they had to wait. Thus, when designing the charging recommendation strategies or optimizing the distribution of charging stations, though all three scenario-related factors matter, the *Waiting Time* may be assigned with lower weight compared to others. However, the exact weights may need to be further tuned based on a larger dataset.

At the same time, we observed inter-relationships among the influential driving experience factors, BEV-related factors, and range-anxiety-related factors, which were absence in our prior work

(Wang et al., 2023). Similar to what has been found in previous studies, we found that BEV driving experience was positively associated with expected "range buffer" (as measured by *Comfort Range 25%*), potentially because experienced drivers are more cautious with trip planning when using BEVs. This result is also in line with the SHAP analysis of BEV driving experience – those who had a higher level of BEV experience were more likely to charge at the upcoming rest area. Additionally, as shown in Table 6, for users who owned a vehicle with shorter display mileages, they would expect a shorter range buffer in a trip (i.e., longer *Comfort Range 50%*) but preferred to charge at the destination according to SHAP analysis. Given that the ranges and distances in the scenarios were all proportional to the *Display Mileage*, it is possible that when using vehicles with shorter ranges, the drivers would experience less uncertainty in the range estimation. This indicates that increasing the maximum BEV range may not always alleviate distance-related range anxiety. However, it should be noted that the maximum BEV range was not associated with users' preferred range buffer when the battery level was already low (e.g., 25% or less), as no association between the *Display Mileage* and the *Comfort Range 25%* has been observed. It is possible that the range anxiety would always be high at critical SoC levels regardless of the maximum BEV range. Finally, we found that age and extraversion were connected with range-anxiety-related factors in BN analysis, while no significant linear relationships were observed in the post-BN regression analysis. Given that age and extraversion can affect charging decisions, it is possible that the variations in age and extraversion will lead to complex nonlinear patterns in range-related anxiety, which can hardly be captured by the linear-regression analysis. Future research may need to better quantify these relationships using non-linear approaches (Motulsky & Ransnas, 1987).

## Limitation

While our study provides valuable insights into distance- and time-related range anxiety, we acknowledge several limitations that may affect the generalizability of our findings. One limitation is the scope of factors considered in our scenarios. We focused exclusively on three scenario factors (i.e., Waiting Time, SoC, and Rest Trip) to highlight the role of time-related range anxiety, given that incorporating additional factors would dramatically increase the number of scenarios required. A larger number of scenarios would fatigue the respondent and potentially compromise the validity of the results. Therefore, future research needs to explore the validity of our conclusions when the influence of additional factors (e.g., charging costs, station availability, and convenience) are considered. Next, from the perspective of charging behavior prediction, neural networks with more complex structures might achieve better performance. In this work, a few selected ML methods were used for behavior modeling as they have good explainability, while future work should explore more advanced deep-based methods with better prediction performance and good explainability. Additionally, though the survey study can collect data from a broad population (therefore increasing the generalizability of the conclusions) and collect more psychological variables compared to research based on behavioral data only, the validity of the results might be compromised by the response biases as what users say may not always match what they do. Therefore, future research should consider integrating the data-driven methods and survey study (e.g., through a naturalistic driving study) to further validate the findings of our study.

## Conclusion

In summary, based on users' charging decisions in scenarios where distance-related and time-related anxiety co-exist, for the first time, we found that both distance and waiting time matter to drivers' charging decisions, and the scenario-related factors interact with each other. Drivers also put less weight on time-related anxiety compared to distance-related anxiety. This finding can guide the design of charging recommendation systems and the optimization of the charging station network. Specifically, both reachability (associated with distance-related range anxiety) and efficiency (associated with time-related range anxiety) should be considered as either objectives or constraints when optimizing the recommendation system or charging station network. Further, we found that users' charging decisions can be moderated by BEV users' psychological states, range anxiety, and their experience with the BEVs, and these influential factors have inter-relationships with each other. As a benefit of the BN-regression mixed approach, we also identified factors that are more closely related to range anxiety – some influential factors may influence charging decisions by indirectly affecting other factors. Thus, on one hand, countermeasures to moderate users' charging decisions may focus more on the factors that are more directly related to charging decisions, i.e., range-anxiety-related factors; on the other hand, future studies may explore more influential factors of range anxiety in order to better model BEV users' charging decisions. Finally, the ML-SHAP-BN-regression approach may be applied to other fields where complex inter-relationships among factors exist (e.g., trust in automation (Huang et al., 2024)) to explore hierarchical structure among the influential factors.